\documentclass[a4paper,11pt]{article}
\pdfoutput=1 

\usepackage{jcappub} 
\usepackage{float}
\usepackage{multirow}
\usepackage[T1]{fontenc} 
\usepackage{subfig}
\usepackage{siunitx}
\def\be{\begin{equation}}
\def\ee{\end{equation}}
\def\barr{\begin{array}}
\def\earr{\end{array}}
\def\bea{\begin{eqnarray}}
\def\eea{\end{eqnarray}}
\def\bfig{\begin{figure}}
\def\efig{\end{figure}}

\title{Cutoff of IceCube Neutrino Spectrum due to t-channel Resonant Absorption by C$\nu$B}
\author[a]{Subhendra Mohanty}
\author[a,b]{Ashish Narang,}
\author[a]{Soumya Sadhukhan}

\affiliation[a]{Physical Research Laboratory, Ahmedabad, 380009, India}
\affiliation[b]{Indian Institute of Technology, Gandhinagar, 382355, India}

\emailAdd{mohanty@prl.res.in}
\emailAdd{ashish@prl.res.in}
\emailAdd{soumyas@prl.res.in}

\abstract{The non-observation of neutrinos by the IceCube at the Glashow resonance energy of 6.3 PeV has been a long standing unresolved issue. In this paper we propose a t-channel neutrino absorption by the C$\nu$B, which causes a cutoff at 4.5 PeV neutrino energy, to explain the IceCube observations. We present a neutrinophilic 2HDM where the neutrino masses are generated by a low scale seesaw mechanism. A $\mathcal{O}$(10) MeV scalar mediates the interactions between left and right handed neutrinos and generates the t-channel diagram used for explaining the absence of Glashow resonance. The same scalar mediates the annihilation of the dark matter and generates the correct relic density.
}

\keywords{IceCube, Cosmic Rays, Glashow resonance, $\nu$2HDM}

\begin{document}
\maketitle
\flushbottom

\section{Introduction}
\label{sec:intro}
IceCube neutrino detector has brought the idea of neutrino astronomy to reality after more than 60 years of its proposal by pushing the energy range of observation and also by combining with the multi-messenger observation. IceCube has observed a total of $82$ high energy cosmic neutrino events. 
After six years of its operation, a clear 6$\sigma$ excess of events is observed at IceCube 
for energies above 60~TeV and these events cannot be explained by the atmospheric 
neutrinos~\cite{Aartsen:2015rwa}. 
The initial choices to explain the ultra high energetic (UHE) neutrino events were different astrophysical sources~\cite{Cholis:2012kq, Anchordoqui:2013dnh, Murase:2014tsa, Sahu:2014fua}. 
As the recent observation of the 290 TeV neutrino~\cite{IceCube:2018dnn, IceCube:2018cha} indicates, exploring these sources and the spectrum of neutrinos observed at IceCube lead us towards the conclusion that the events do point back to clear identifiable single power law astrophysical sources \cite{Meszaros:2017fcs}(i.e. Active Galactic Nuclei (AGN), Gamma Ray Bursts (GRB) etc), mainly pointing to the neutrinos from the blazars. 

But some puzzles remain, the entire spectrum of IceCube events from 60 TeV to 10 PeV cannot 
be explained by a single power law of neutrino flux like $\Phi_\nu= \Phi_0 E_\nu^{-\gamma}$\cite{Chen:2014gxa, Anchordoqui:2014hua, Anchordoqui:2016ewn, Palladino:2018evm, Sui:2018bbh}. A power law flux predicts the presence of a Glashow resonance at energy $6.3$~PeV. Still no Glashow resonance, i.e., an excess of events at $6.3$~PeV has been observed at IceCube till now~\cite{Palomares-Ruiz:2015mka}. 

If a single power law astrophysical neutrino flux is used to explain the absence of IceCube events at higher energy bins $E_{\nu}> 3 $~PeV, then either the initial flux amplitude has to be very low or 
the spectrum has to be a steep one. In both the cases it is difficult to have a good fit of the
low energy bins ($60-600$~TeV) and also the two super-PeV bins ($< 2$~PeV). 
Decrease of initial flux amplitude can fit neutrino events at some bin, but the predicted events 
at other energies show huge mismatch with the observation.  
If a steeper neutrino energy spectrum is taken, IceCube event distribution in the sub-PeV bins can be explained by fixing a proper neutrino flux amplitude. 
The steepness of the spectrum will result in deficiency of predicted events at energies $1-3$~PeV.
If we want a fit at lower energy ($\sim 100$~TeV) bins, a less steeper flux predicts an excess of neutrino events at higher energy bins, but IceCube has not observed that effect till now. 

Various explanations of the observed 1 PeV excess feature in the IceCube event spectrum include neutrinos resulting from PeV dark matter decay or annihilation~\cite{Esmaili:2013gha, Murase:2015gea, Dev:2016qbd, Bhattacharya:2014yha,Borah:2017xgm}, the resonant production of leptoquarks~\cite{Barger:2013pla, Dutta:2015dka, Dey:2015eaa, Chauhan:2017ndd,Dey:2017ede} and the interactions involving R-parity violating supersymmetry \cite{Dev:2016uxj}. On the other hand there are depletion models which try to explain the non-observation of the Glashow resonance~\cite{PhysRev.118.316,Berezinsky:1977sf,Berezinsky:1981bt,Sahu:2016qet,Kistler:2016ask} neutrinos produced when a real $W^-$ is produced by the process ${\bar \nu}_e e^- \to W^-$. The decay of the real $W$ is expected to give hadron and lepton shower or lepton track events~\cite{Bhattacharya:2011qu}. Depletion of high-energy neutrinos can occur via oscillation to sterile neutrinos in pseudo-Dirac neutrinos~\cite{Joshipura:2013yba} and for visible decay~\cite{Pakvasa:2012db}. Exotic scenarios have also been invoked to explain a cutoff at the Glashow resonance energies such as Lorentz violation~\cite{Tomar:2015fha, Anchordoqui:2014hua} and CPT violation~\cite{Liao:2017yuy}. 
 
We explore a new phenomenon in the context of IceCube observation here. We discuss a scenario 
where an ultra high energetic (UHE) neutrino originates from an astrophysical source and while it travels towards the IceCube detector at the Earth, it interacts with the cosmic neutrino background 
(C$\nu$B) and get absorbed through a t-channel (and also u-channel) process mediated by a scalar causing a suppression of the neutrino flux. Earlier this kind of absorption process is explored in context of the production of a s-channel resonance ~\cite{weiler1999,farzan,Araki:2014ona,Araki:2015mya,Chauhan:2018dkd}, and also in the context of secret neutrino interactions~\cite{Ng:2014pca,Ioka:2014kca,Blum:2014ewa}. Here we present a different scenario where we have an energy cutoff where the absorption process kicks off and it depends on kinematic viability. Then we explain the presence of a peak from the t-channel resonance, that causes sharp dip in the IceCube spectrum. When the absorption process is on it vanishes the incoming neutrinos which results in a wash out of neutrino events at IceCube. Therefore, fixing the cutoff energy, the cross section and the energy where it peaks one can explain the disappearance of the Glashow resonance peak at around $E_{\nu} \sim 6.3$~PeV.  

In order to explain the features of IceCube event spectrum, we propose a variant of two Higgs doublet model (2HDM) where only the second doublet couples to right handed neutrinos introduced in this model. This doublet has a tiny vacuum expectation value ($v_2 \sim$ keV), which mixes the left and right handed neutrinos and therefore, provides a seesaw mechanism. We term this seesaw mechanism through the neutrinophilic scalar doublet as 'neutrinophilic seesaw', that gives a $0.1~$eV scale neutrino mass with presence of $10$~MeV scale right handed neutrinos. In this model one high energetic cosmic $\nu_L$ interacts with another background $\nu_L$ through a t-channel diagram with a light CP-even scalar from the second doublet, to produce two right handed neutrinos that the IceCube cannot detect. Once the cosmic ray neutrinos (CR$\nu$) interact through this process, they get absorbed to create a dip in the incoming neutrino flux which subsequently shows a dip in the IceCube. The cosmic neutrino hits three background neutrino mass eigenstates and gets absorbed which results in three dips in the neutrino flux spectrum at three different incoming energies. The absence of Glashow resonance can be explained by the first dip at around $E_{\nu} \sim 6.3$~PeV.

This paper is organized as follows. In Sec.~\ref{sec:glashow} we have explained the nature of the observed IceCube spectrum and the method to theoretically compute the event spectrum where we notice the absence of Glashow resonance in that spectrum. 
In the next Sec.~\ref{sec:capnu}, we explain the absence of neutrino flux 
at PeV scale through absorption in the cosmic neutrino background and the possible flux modification due to that, in a model-independent way.
In the next Sec.~\ref{sec:model}, we discuss a $\nu$2HDM model re-casted in a way where second Higgs doublet can be used to have a seesaw mechanism through $Z_2$ breaking and that can explain the tiny neutrino mass. We also discuss possible collider and flavor constraints on this model. 
In Sec.~\ref{sec:tabsorp}, we show how a t-channel absorption of CR$\nu$ through a ultra light scalar propagator takes place in this model and how the cross section gets modified with the model parameters. Then we show how this process with a threshold energy 4.5 PeV leads to depletion of events at 4.5-10 PeV energy bins. In Sec. \ref{sec:singdm} we extend the model to include a singlet scalar dark matter, which satisfies the relic density constraints along with small scale constraints on self-interacting dark matter. Finally, we summarize and conclude in Sec.~\ref{sec:conc}.
\section{Absence of Glashow Resonance at IceCube}
\label{sec:glashow}
\begin{figure*}[h!]
\begin{center}
\includegraphics[width=5.2in,height=2.4in,angle=0]{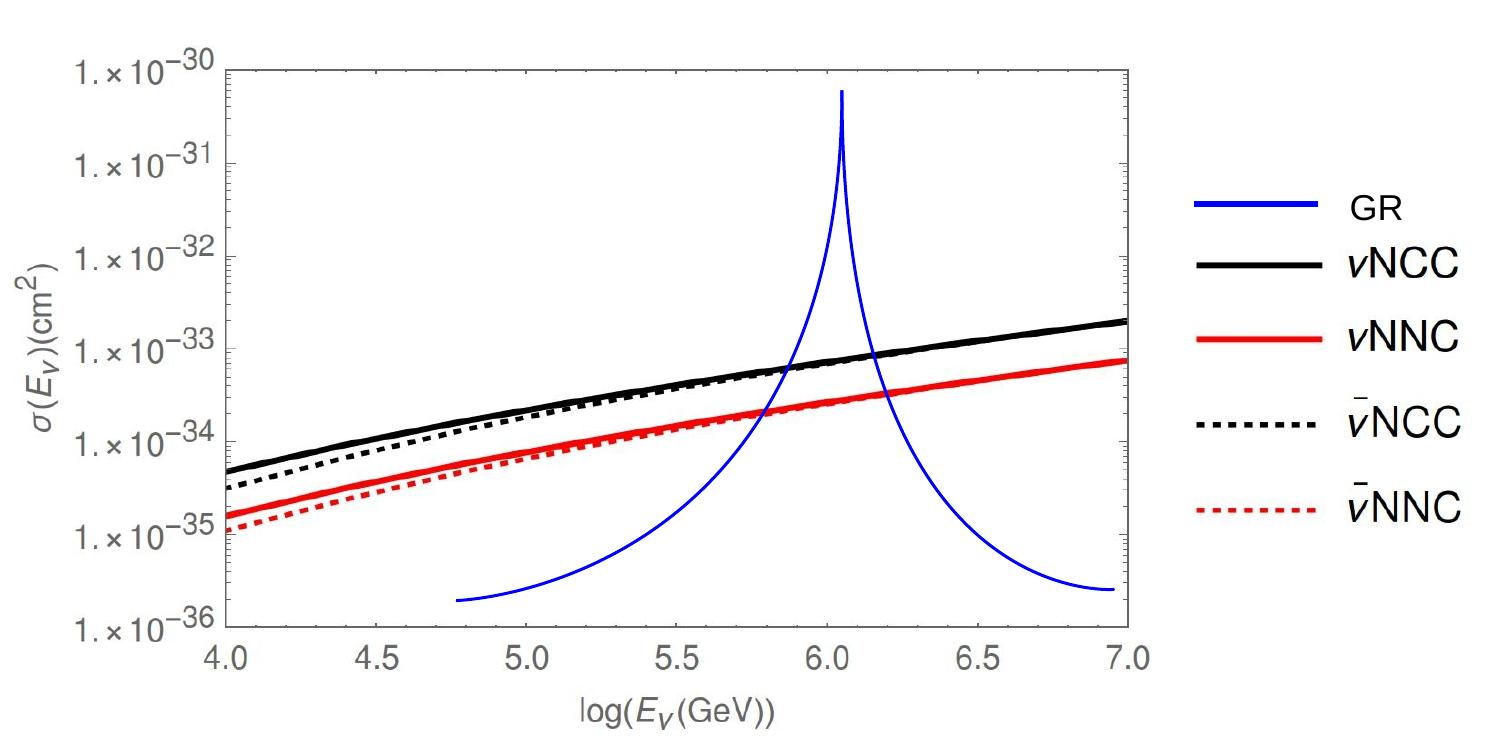}~~
\caption{The cross section of $\nu$-nucleon and $\bar{\nu}$-neucleon interaction along with the $\bar{\nu}_{e} e^{-}$ responsible for GR are shown here. The CS of $\bar{\nu}_{e} e^{-}$ is very large compared to interaction with nucleons at energy around 6.3 PeV  }\label{fig:smcs}
\end{center}
\end{figure*}

\begin{figure}[H]
\begin{center}
\includegraphics[width=5.2in,height=3.2in,angle=0]{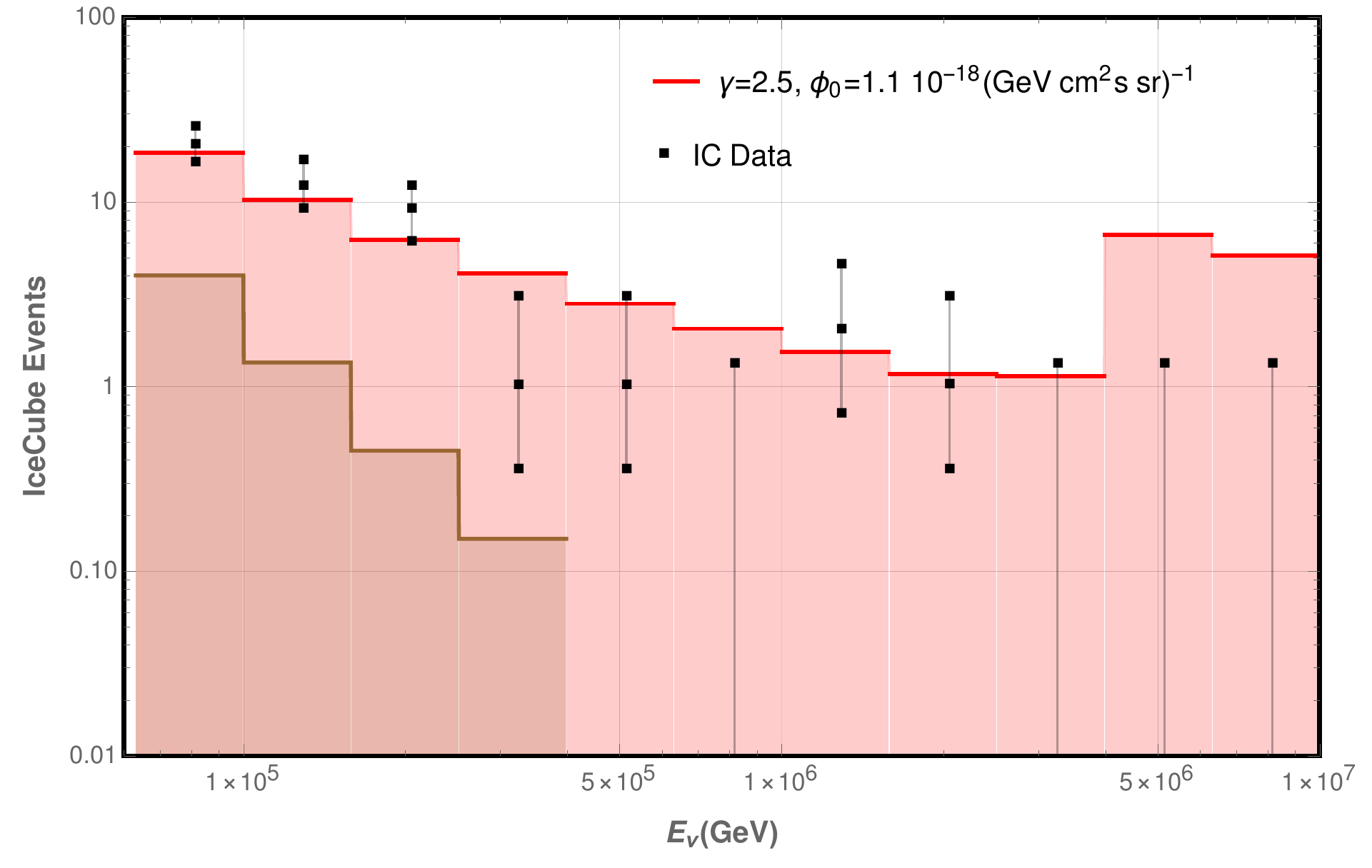}~~
\caption{The IceCube event spectrum with a single power law flux of the cosmic ray neutrinos is shown here. The figure shows that with SM interactions and a single power law flux one cannot fit the IceCube data completely. SM with single power law flux expects events in the last three bins due to GR but there are no events seen in these bins.}\label{fig:wocap}
\end{center}
\end{figure}

The high energy cosmic ray neutrinos are detected at IceCube due to their deep inelastic scattering with the quarks and electrons present in the detector volume. The interactions of UHE neutrinos with electrons in the detection volume are proportional to the electron mass and therefore have negligible interaction rate compared to the neutrino-nucleon interactions. However, the resonant scattering of electron anti-neutrino
\begin{equation}
\bar{\nu}_{e} e^{-} \longrightarrow W^{-}  \longrightarrow \rm{hadrons+leptons}
\end{equation}
with energy $E_\nu=M_{W}^{2}/2 m_{e}\simeq 6.3$ PeV has an enhanced probability of interaction with the atomic electrons in the ice to produce the on-shell $W^{-}$ boson. This is the so-called Glashow resonance (GR)~\cite{PhysRev.118.316}. The expressions for the differential cross section for these interactions can be found in ~\cite{rgandhi}. The cross section at GR is about 300 times higher than that of the charged current (CC) neutrino-nucleon interaction, see Figure~\ref{fig:smcs}. 
As a result of higher cross section of $\bar{\nu}_{e} e^{-}$ interaction at energy $6.3$~PeV there are events expected in the IceCube event spectrum at that energy. However, as shown in figure~\ref{fig:wocap}, in the 6 years of its data collection IceCube has observed no GR events. This is in general referred to as the "missing Glashow resonance" problem.

The absence of Glashow resonance in the IceCube event spectrum and the observation of more numbers of PeV events than expected have been the major outcome of IceCube neutrino detection. 
The number of events at IceCube in the deposited energy interval ($E_{i},E_{f}$) is given by\cite{Chen:2013dza, Anchordoqui:2014hua, Palomares-Ruiz:2015mka}

\begin{equation}
\label{noe}
\mathcal{N} =T ~N_A~ \int_{0}^{1}dy \int_{E_{\nu}^{ch}(E_i, y)}^{E_{\nu}^{ch}(E_f, y)}dE_\nu ~ \mathcal{V}_{eff}(E_{dep}^{ch})~ \Omega(E_\nu)~ \dfrac{d\phi}{dE_\nu} \dfrac{d\sigma}{dy}^{ch}.
\end{equation}
where the total exposure time $T = 2078$ days, $N_A = 6.023\times 10^{23}$ cm$^{-3}~$ water equivalent is the Avogadro's Number, and $ch$ denotes the interaction channel (neutral current (NC), charged current (CC)). $E_{dep}^{ch}$ is the deposited energy as explained in ~\cite{mileo}. We have used $\Omega=2\pi$ and  $\Omega=4\pi$ for the super-PeV ultra high energetic bins and sub-PeV IceCube bins respectively. The terms appearing in the above expression are explained in detail below.
\begin{itemize}
\item $\dfrac{d\phi}{dE_\nu}$ is the flux of the cosmic ray neutrinos. It is assumed that for neutrinos and anti-neutrinos of each flavour the flux is isotropic and is given by a power law flux parametrized as
\begin{equation}
	\label{flux}
	\dfrac{d \Phi}{dE_\nu} = \phi_0 \left(\frac{E_\nu}{100~\text{TeV}} \right)^{-\gamma}.
	\end{equation}
We take
	\be \phi_0 = 1.1 \times 10^{-18} \text{GeV}^{-1} s^{-1} sr^{-1} cm^{-2} \ee
	 \be \gamma = 2.5. \ee

\item $\mathcal{V}_{eff}(E_{dep}^{ch})$ is the effective volume of the detector available for the interaction, given by 

\begin{equation}
\mathcal{V}_{\textrm{eff}}(x) = \begin{cases}
\dfrac{1 + d \, x^q}{c \, x^q} & \textrm{if  } x \geq 0 \\[3ex]
0 &  \textrm{if  } x < 0 ~,
\end{cases}
\end{equation}

where $x \equiv \log_{10}\left(\frac{E^{ch}_{dep}}{E_{\textrm{th}}}\right)$ with $E_{\textrm{th}}=10$ TeV.

\item $\dfrac{d\sigma}{dy}^{ch}$ is the SM differential cross section of the neutrino-nucleon interaction. Depending on the channel $ch$ of the interaction, these are given as 
\begin{equation}
\frac{d^2\sigma}{dx\,dy}^{\!\!\!(CC)}\!\!\!=\,\frac{G^2_F}{\pi}\frac{2M^4_W}{(Q^2+M^2_W)^2}M_NE_{\nu} \,\{xq(x,Q^2)+ x\bar{q}(x,Q^2)(1-y)^2\},
\label{cc}
\end{equation}

\begin{equation}
\frac{d^2\sigma}{dx\,dy}^{\!\!\!(NC)}\!\!\!=\frac{G^2_F}{2\pi}\frac{M^4_Z}{(Q^2+M^2_Z)^2}M_NE_{\nu}\,\{xq^0(x,Q^2)+x\bar{q}^0(x,Q^2)(1-y)^2\}
\label{nc}
\end{equation}
and for the processes contributing to the Glashow resonance
\begin{eqnarray}
	\lefteqn{\frac{d\sigma(\bar{\nu}_e e \rightarrow \bar{\nu}_e e)}{dy} =
	\frac{G_F^2 m_{e} E_\nu}{2\pi} \left[ \frac{R_e^2}{\left(1+2 m_{e} E_\nu
	y/M_Z^2\right)^{\!2}}\; + \right.}  \nonumber \\ & & \hspace{0.4in}\left.
	\left|\frac{L_e}{1+2 m_{e} E_\nu y/M_Z^2} + \frac{2}{1-2mE_\nu/M_W^2 + i
	\Gamma_W/M_W}\right|^2(1-y)^2
	\right]\!,
	\label{nubaree}
\end{eqnarray}
\begin{equation}
	\frac{d\sigma(\bar{\nu}_e e \rightarrow \bar{\nu}_\mu \mu)}{dy}	=
	\frac{G_F^2 m_{e} E_\nu}{2\pi}\frac{4(1-y)^2[1-(m_{\mu}^2-m_{e}^2)/2 m_{e}E_{\nu}]^2}
	{(1-2 m_{e}E_\nu/M_W^2)^2+\Gamma_W^2/M_W^2}\; ,
	\label{muviaW}
\end{equation} and
\begin{equation}
	\frac{d\sigma(\bar{\nu}_e e \rightarrow \hbox{ hadrons})}{dy} =
	\frac{d\sigma(\bar{\nu}_e e \rightarrow \bar{\nu}_\mu \mu)}{dy}
	 \cdot \frac{\Gamma(W\rightarrow \hbox{ hadrons}) }
	 {\Gamma(W \rightarrow \mu \bar{\nu}_\mu)}\; ,
	\label{Wtohads}
\end{equation}
where $\Gamma_{W}=2.09$ GeV is the decay width of the W boson, $L_e=2 Sin^{2}\theta_W-1$ and $R_e=2 Sin^{2}\theta_W$ are chiral couplings of Z to electron, and $M_W$ and $M_N$ are the $W$ boson and the nucleon masses respectively, $-Q^2$ is the invariant momentum transferred to hadrons, and $G_F$ is the Fermi constant. The Bjorken scaling variable $x$ and the inelasticity $y$ are defined as
\begin{equation}
x= \frac{Q^2}{2M_NE_{\nu}y}\qquad\mbox{ and }\qquad y=\frac{E_{\nu}-E_{\ell}}{E_{\nu}},
\end{equation}
where $E_{\nu}$ is the energy of the incoming neutrino and $E_{\ell}$ is the energy carried by the outgoing lepton in the laboratory frame. $q(x,Q^2)$, $\bar{q}(x,Q^2)$, $q^0(x,Q^2)$ and $\bar{q}^0(x,Q^2)$ are quark distribution functions in the nucleon, the expression of and further details on which can be found in ~\cite{rgandhi,mileo}. 
Figure \ref{fig:wocap} shows the IceCube event spectrum with 6 years IceCube data. We can see that no events have been seen where the dramatic increase of the event rate was expected due to the Glashow resonance.
 
\end{itemize} 
\section{Cosmic Ray $\nu$ Absorption by C$\nu$B}
\label{sec:capnu}
\begin{figure}[H]
\begin{center}
\includegraphics[width=3.2in,height=2.0in,angle=0]{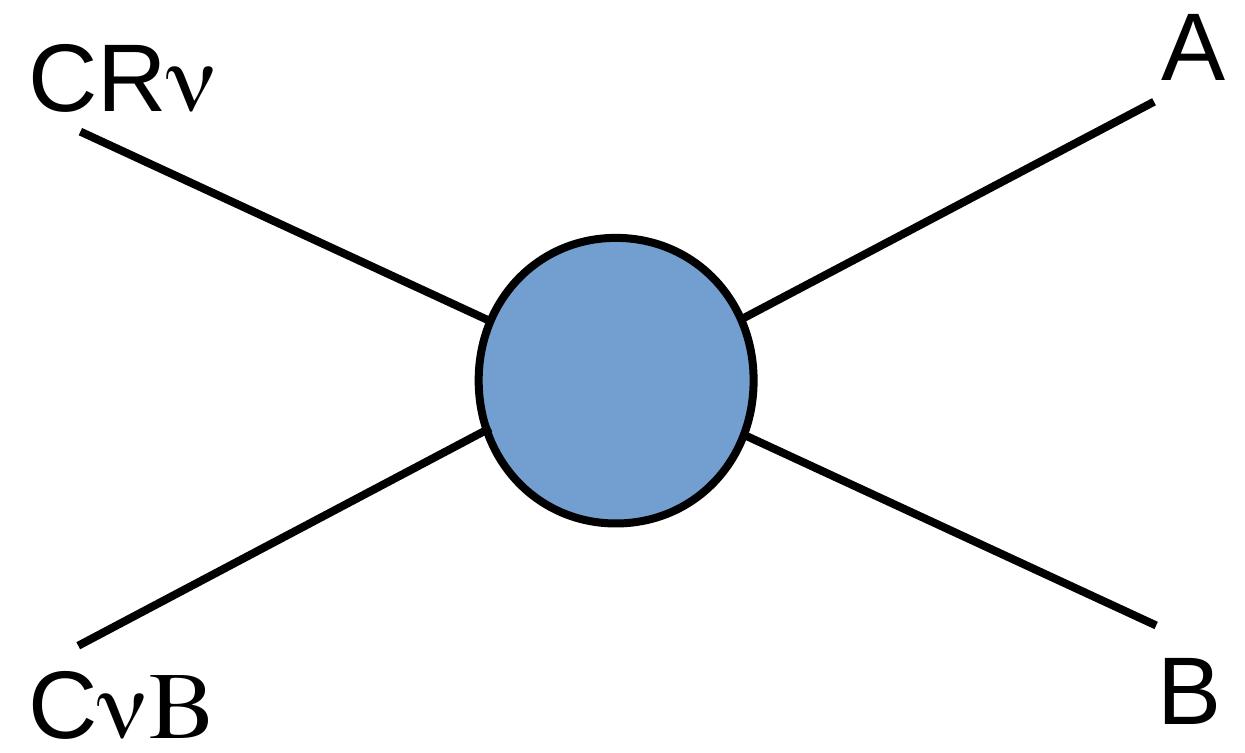}~~
\caption{Schematic diagram of CR$\nu$ capture by C$\nu$B neutrino. }\label{fig:cap}
\end{center}
\end{figure}

Cosmic neutrino background (C$\nu$B) is the relic of the hot plasma from the early universe. These neutrinos decouple from the hot plasma at 1 MeV and now expected as an isotropic background at temperature $T_\nu=1.95$K.
To resolve the enigma of absence of Glashow resonance at IceCube, we propose a scenario where CR$\nu$s interact with the C$\nu$B neutrinos inelastically and produce new particles which cannot produce any signatures at IceCube. This scenario is shown schematically in figure~\ref{fig:cap}. When the CR neutrinos of energy much greater than $T_\nu$ hit the C$\nu$B neutrinos, C$\nu$B neutrinos can be considered as stationary in that process in the laboratory frame. Since this absorption process would take place only if the CR$\nu$ has enough energy to produce particles in the final state (particles A, B in figure \ref{fig:cap}), therefore depending on the masses of the particles in the final state, the CR$\nu$ of different energies can be absorbed. The absorption of CR$\nu$ by C$\nu$B through production of a new particle has been discussed earlier~\cite{weiler1982,weiler1999,yoshida1994,yoshida1997,farzan,Ibe:2014pja}.
In this paper we explore the t-channel resonant absorption of the CR$\nu$ by C$\nu$B. Unlike s-channel absorption, the mass of the mediator does not decide CR$\nu$ of which energy will be absorbed. The onset of the process is controlled by the masses of the particles in the final state and the C$\nu$B mass. Depending on these masses of the final state particles and the targets, the t-channel absorption has a threshold, i.e., the process takes place only if the CR$\nu$ has sufficient energy. Once the CR$\nu$ has enough energy for the process to happen then the strength of the interaction is decided by other parameters like coupling and the mediator mass. 
The absorption of CR$\nu$ produces a dip in the CR$\nu$ flux starting at the threshold energy of the t-channel process. The mean free path is, 
\begin{equation}
\label{mfp}
\lambda_i (E_i, z) =  \left( \sum_{j} \int \frac{d^3 \mathbf{p}}{(2 \pi)^3} f_{j} ( p ,z) \sigma_{ij}( p, E_i, z) \right)^{-1} \approx  \left( n_\nu(z) \sum_{j} \sigma_{ij}( p, E_i, z) \right)^{-1}
\end{equation}
where $f_i$ is the distribution function for the neutrinos given by, 
\begin{equation}
f_i (p, z) ^{-1} = \exp \left( \frac{p}{T_{i}(1+z)} \right) + 1 
\end{equation}
and $T_{i} = 1.95~K$ for all three mass states. Away from the sources, due to the mixing, flavor ratio of neutrino in the cosmic ray flux is (1:1:1).  As will be discussed in the Sec. \ref{sec:tabsorp}, the mean free path (MFP) of the CR$\nu$ is much greater than $\mathcal{O}$(1) Mpc, i.e., the coherence length. The coherence is lost after traveling such a long distance, and therefore the scattering process can be described in terms of mass eigenstates. Also, the (1:1:1) flavor ratio directly translates to (1:1:1) ratio in the three mass eigenstates. Away from the sources, we assume a power law flux for each flavor, in turn for each mass eigenstate of neutrino. We express the modified flux due to the absorption as
\begin{equation}
\left(\dfrac{d \Phi}{dE_\nu}\right)_{cap} = \exp \left[   - \int_{0}^{z_s} \frac{1}{\lambda_i} \frac{dL}{dz} dz\right]\dfrac{d \Phi}{dE_\nu}
\end{equation}
\label{eq:capflux}
where $z_s$ denotes the redshift of source and,
\begin{equation}
\frac{dL}{dz} = \frac{c}{H_0 \sqrt{ \Omega_m (1 + z)^3 + \Omega_\Lambda}}. 
\end{equation}

The details of the t-channel absorption are given in Sec.~\ref{sec:tabsorp} in the context of $\nu$2HDM theory.
\section{The $\nu \rm 2HDM$: Neutrinophilic Seesaw}
\label{sec:model}
The $\nu$2HDM theory~\cite{Gabriel:2006ns, Machado:2015sha} allows us to have the t-channel absorption explained in the Sec.~\ref{sec:capnu}. In this section we explain the $\nu$2HDM theory which is based on the symmetry group $SU(3)_c \times SU(2)_L \times U((1)_Y \times Z_2$. 
We have three EW singlet right-handed (RH) neutrinos, $N_{R_{i}}$, for each flavor of SM lepton. In addition to that the model has two Higgs doublets, $\Phi_{1}$ and $\Phi_{2}$. All the charged fermions and the Higgs doublet $\Phi_{1}$, are even under the discrete symmetry, $Z_2$, while we give charge to the RH neutrino and the Higgs doublet $\Phi_{2}$ under $Z_2$. Such a setup leads to Yukawa structure in which all the charged fermions couple with $\Phi_{1}$ only and the left-handed neutrinos, together with the right-handed neutrino added here, couple to the Higgs doublet $\Phi_{2}$. 
We break the discrete symmetry $Z_2$ by a vev of $\Phi_{2}$, and we take $v_2 = \langle\Phi_{2}\rangle \sim 0.1$~keV. 
As a result of the spontaneous breaking of $Z_2$ a seesaw mechanism will be initiated and the diagonal of neutrino mass matrix will appear through interaction with $\Phi_{2}$.

The Higgs sector is considered to be CP invariant here. 
The 2 Higgs potential with $Z_2$ symmetry is given as ~\cite{Machado:2015sha}
\begin{align}
\ V =
-\mu^2_1~\Phi_{1}^{\dag}\Phi_{1}-\mu^2_2~\Phi_{2}^{\dag}\Phi_{2}+ m_{12}^2~(\Phi_{1}^{\dag}\Phi_{2}+h.c.)+ \lambda_1(\Phi_{1}^{\dag}\Phi_{1})^{2}+\lambda_2(\Phi_{2}^{\dag}\Phi_{2})^{2}+ 
\lambda_3(\Phi_{1}^{\dag}\Phi_{1})(\Phi_{2}^{\dag}\Phi_{2})+ \nonumber\\
\lambda_4|\Phi_{1}^{\dag}\Phi_{2}|^{2} + \frac{1}{2}\lambda_5[(\Phi_{1}^{\dag}\Phi_{2})^{2}+(\Phi_{2}^{\dag}\Phi_{1})^{2}].
\label{pot}
\end{align}
In the above potential, $\lambda_{6,7}=0$.
After the EW symmetry breaking, the two doublets can be written as follows in the unitary gauge
\begin{align}
\Phi_{1} =  \frac{1}{\sqrt{2}} \left(\begin{array}{c}
                                                 \sqrt{2} (v_2/v)H^{+} \\
                                                 h_0 + i (v_2/v)A +v_1 \\
                                               \end{array}
                                             \right),    \nonumber
\end{align}
\begin{align}
           \Phi_{2} = \frac{1}{\sqrt{2}} \left(
                                               \begin{array}{c}
                                                 -\sqrt{2} (v_1/v) H^{+} \\
                                                 H_0 - i (v_1/v) A + v_2\\
                                               \end{array}
                                             \right),
 \end{align}
where charged fields $H^{\pm}$, two neutral CP even scalar fields $h$ and $H$, and a neutral CP odd field $A$ are the physical Higgs fields and $v_1 = \langle\Phi_{1}\rangle$, $v_2 = \langle\Phi_{2}\rangle$, and $v^{2} = v^{2}_1 + v^{2}_2$. 
There is an orthogonal mixing of the charged and CP odd interaction states with corresponding charged and neutral Goldstone modes with a mixing angle $\beta$.
As a result of the mixing mass eigenstates $H^{\pm}, A$ and massless Goldstone bosons are produced. 
The mixing angle is expressed as $\tan \beta = \frac{v_2}{v_1}$. 
The masses charged Higgs and the CP-odd Higgs in this model are of the order 100 GeV. This model also gives rise to a very light scalar $H$ with mass varying from 1 eV to 1 GeV. 
%


%
The mass eigenstates $h, H$ are related to the weak eigenstates $h_0, H_0$ by
\begin{align}
\label{mix}
 h_0 = c_{\alpha} h + s_{\alpha} H, ~H_0 = -s_{\alpha} h + c_{\alpha} H,
  \end{align}
where $c_{\alpha} = \cos\alpha, s_{\alpha} = \sin\alpha$, and
are given by
\begin{align}
\label{scmix}
 c_{\alpha} &= 1+O(v^{2}_2/v^{2}_1),\nonumber \\
s_{\alpha} &= -\frac{\lambda_3-\lambda_4-\lambda_5}{2\lambda_1}(v_2/v_1)+O(v^{2}_2/v^{2}_1).
\end{align}
We take $v_{2} \sim 1$~keV (to fit the neutrino mass) and $v_{1} \sim 246$ GeV which results in very small mixing and therefore can be neglected. This results in very small $\tan\alpha$ which in turn gives a small $\tan\beta$. 
Hence, the neutral scalar $h$ effectively behaves like the SM Higgs. 
So we expect that the all the constraints on the coupling of the SM Higgs are also satisfied by $h$, except in the loops.
The effect of $H^{\pm}$ loop in the $h\gamma\gamma$ constraints is given in Ref.~\cite{Seto:2015rma} and also how a sizable Higgs invisible decay is allowed is also discussed.
With no lepton number conservation imposed, $\nu$2HDM allows Majorana mass generation of neutrinos with a low scale seesaw mechanism.

The added three right handed neurinos are gauge singlet Majorana neutrinos $N_{R,\beta}$, all of which transform as odd under the $Z_2$ symmetry. With all the SM fermions being $Z_2$ invariant, the Yukawa interaction in this model in the flavor basis takes the form,
\begin{equation}
\mathcal{L}_{Y}= Y^{d}_{\alpha \beta}\bar{Q}_{L,\alpha} \Phi_{1} d_{R,\beta} + Y^{u}_{\alpha \beta}\bar{Q}_{L,\alpha} \tilde{\Phi}_{1} u_{R,\beta} + Y^{l}_{\alpha \beta} \bar{L}_{L,\alpha}\Phi_{1}l_{R,\beta} + Y^{\nu}_{\alpha \beta} \bar{L}_{L,\alpha} \tilde{\Phi}_{2} N_{R,\beta}+ \mathrm{h.c.}
\end{equation}
If we restrict our model to only one right handed Majorana neutrino $N_{R}$, then the relevant Yukawa and mass terms of the right handed neutrino in the mass basis of the SM neutrinos are written as,
\begin{equation}
\mathcal{L}=y_i \bar{L}_{i} \tilde{\Phi}_{2} N_R+ \frac{m_R}{2} N_R N_R.
\end{equation}
Here the Yukawa couplings $y_i$ are mixture of flavor basis Yukawas $Y_{\alpha \beta}$ for one particular right handed Majorana neutrino.
The neutrino mass matrix takes the form:
\begin{equation}
\label{Higgs}
M_{\nu_i} 	=
	\begin{pmatrix}
		0 \ \  \frac{y_i v_2}{\sqrt{2}}\\  \frac{y_i v_2}{\sqrt{2}} \ \  \frac{m_R}{2}
	\end{pmatrix}.
\end{equation}
Diagonalizing this matrix we compute Majorana neutrino mass term as:
\begin{equation} 
\mathcal{L}_{\nu_{i}} = m_{\nu_i} \nu_{iL} \nu_{iL}  \nonumber 
\end{equation}
with 
\begin{equation}
m_{\nu_i} = \frac{y_i^2 v_{2}^{2}}{m_R}.
\end{equation}
With a Yukawa coupling $y_i \sim O(0.1)$, we get the Majorana neutrino mass of $0.1$~eV for $v_2 \approx 10$~keV with right handed neutrino mass $m_R \sim 10$~MeV. This type of low scale seesaw mechanism was first proposed in the Ref.~\cite{Ma:2000cc}.

\begin{figure}[H]
\begin{center}
\includegraphics[width=4.2in,height=1.7in,angle=0]{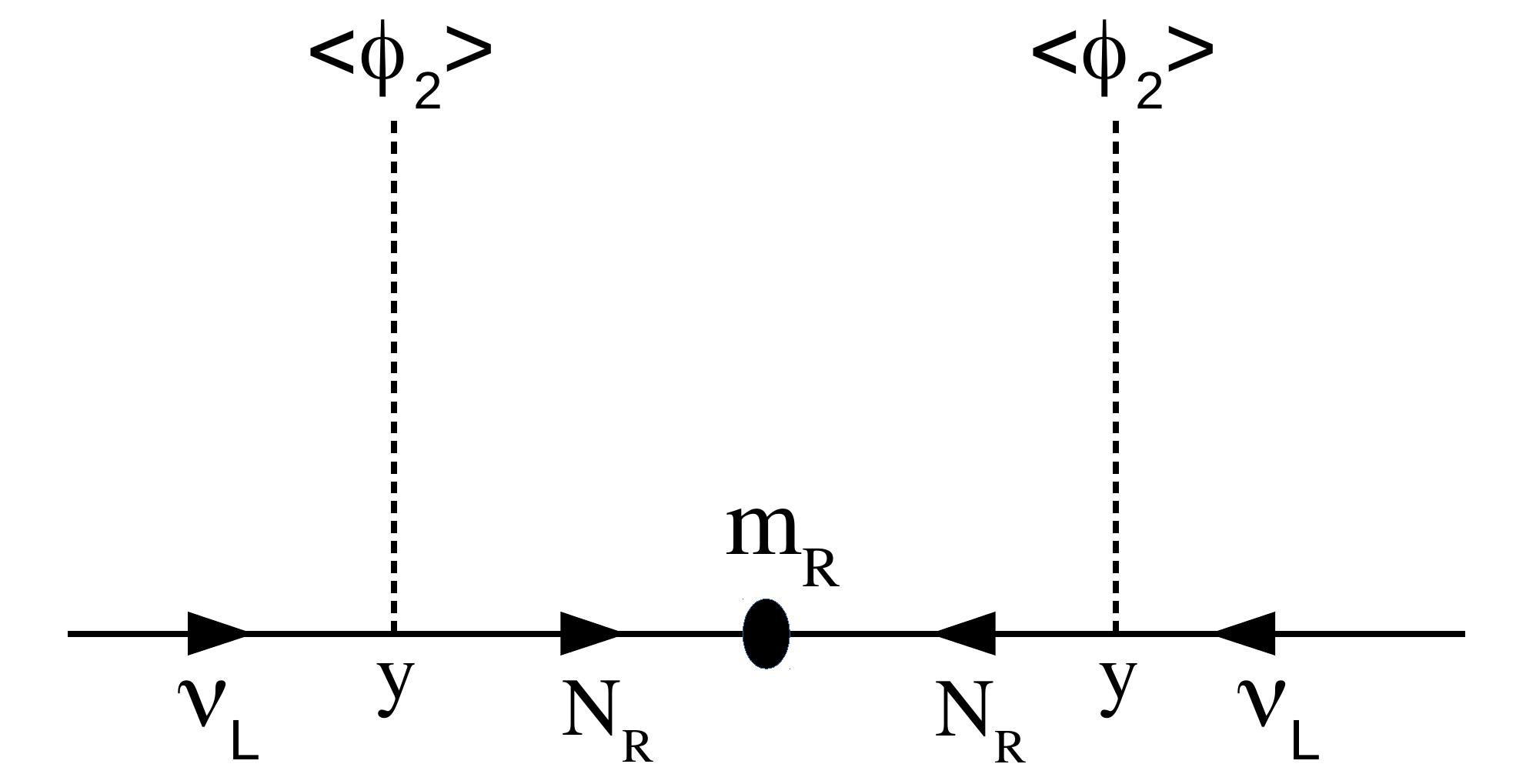}~~
\caption{Majorana neutrino mass generation through seesaw mechanism.}\label{fig:seesaw}
\end{center}
\end{figure}

\subsection{Constraints on the model}
Peskin-Takeuchi oblique parameters S,T, U are measure of corrections to the gauge boson two point functions ($\Pi_{VV}$) \cite{PhysRevD.45.R729}. The deviation of these oblique parameters from SM measured by the LHC are:
\begin{equation}
  \begin{aligned}
    & \Delta S^{SM} = 0.05\pm0.11,\\
    & \Delta T^{SM} = 0.09\pm0.13,\\
    & \Delta U^{SM} = 0.01\pm0.11,\\
  \end{aligned}
\end{equation}
In the model discussed here, we have very tightly constrained scalar sector, since we take $v_2 \ll v_1$ the mixing between the two Higgs is negligible. $\lambda_1$ is fixed by taking $h$ to be the 125~GeV Higgs discovered at LHC. The other CP-even Higgs is very light and for typical quartic couplings within the perturbative limit, the masses of charged scalars and the CP-odd scalar are below the TeV scale. As a result of the presence of a light neutral scalar the oblique parameters S and T will play a decisive role in constraining the model. These constraints are discussed in ~\cite{Sadhukhan:2018nsk}.

The charged Higgs production at LHC in $\nu$2HDM is same as in 2HDM. Due to the smallness of the mixing between the two Higgs the decays of charged Higgs to quarks are highly suppressed by the mixing factor $\tan \beta$. So the dominant channel of charged scalar decay is $H^{\pm}\longrightarrow l^{\pm} \nu$ which is constrained by LEP as $m_{H^{\pm}}$> 80 GeV. Charged Higgs can also contribute to diphoton production through loop. Even with that contribution the diphoton bound is satisfied. The status of constraints from flavor physics is given in \cite{Bertuzzo:2015ada}. Astrophysical consequences of $\nu$2HDM are discussed in~\cite{Sher:2011mx}.

Due to difference in vacuum expectation value of two scalar doublets ($v_2 \ll v_1$), the mixing between two doublets is tiny. As neutrinos in our model couple only to $\Phi_2$, so the SM Higgs coupling to  $\nu + N$ is negligibly small and therefore, that decay does not affect any constraints. The other CP even scalar dominantly from $\Phi_2$ is very light $m_H \sim 10~$MeV and with right handed neutrinos with masses around $15~$MeV the $H \to \nu N$ decay is negligible even with order one scalar Yukawa coupling.
\section{t-channel resonant absorption: Dip in IceCube Spectrum}
\label{sec:tabsorp}
\begin{figure}[H]
\begin{center}
\includegraphics[width=2.3in,height=2.3in,angle=0]{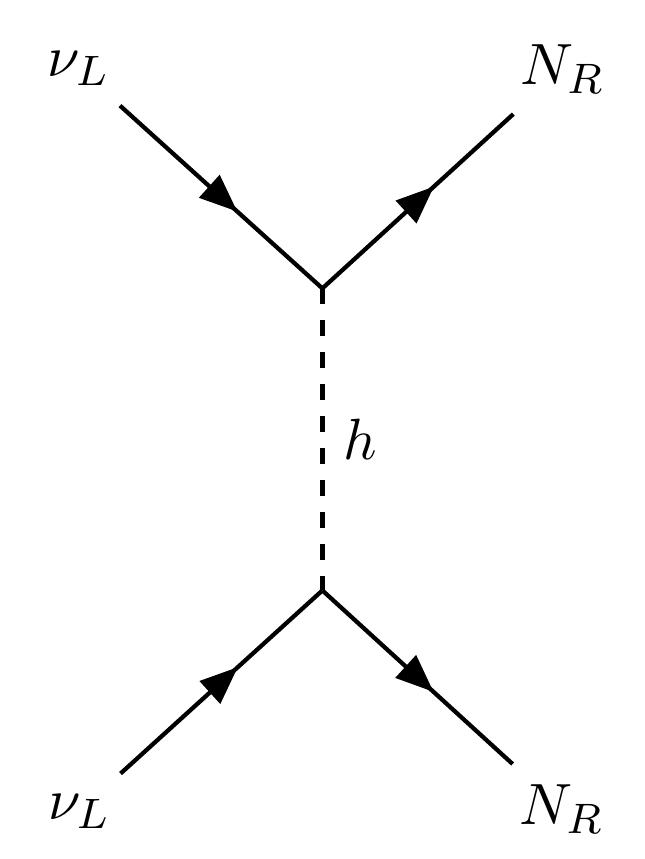}~~
\caption{t-channel absorption of UHE neutrino by C$\nu$B}
\label{fig:tchan}
\end{center}
\end{figure}

In the neutrinophilic model discussed in Sec.~\ref{sec:model}, we have a t-channel diagram shown in Fig. \ref{fig:tchan}. Occurrence of a resonant absorption dip 
is not possible in this model, as there are no right handed neutrinos present 
in the cosmic neutrino background (C$\nu$B), which can absorb an ultra high energy (UHE)
(PeV scale) astrophysical neutrino to produce a MeV-scale scalar resonance. Here through the 
t-channel process UHE neutrino gets absorbed by the neutrino background and releases two 
right handed neutrinos. 
These two right handed neutrinos, unlike their left handed partners, 
do not have charged current and neutral current interaction with IceCube matter. 
Therefore those will not be detected in the IceCube, which results in vanishing 
one astrophysical neutrino in this process. 

The t-channel diagram cross matrix element is computed as: 
\be
M^2= \frac{4 y_i^2 y_j^2}{(t - m_h^2)^2} \left(-\frac{1}{2}(t-m_R^2) + m_{\nu_{i}} m_R \right)^2
\ee
where t represents the energy transfer to the final state right handed neutrinos. Here $m_h$ and $m_R$ are the ultralight scalar mass and the 
right handed neutrino mass respectively, with y being the neutrino-scalar Yukawa coupling. 
Depending on the mass of the final state right handed neutrinos $N_R$ the t-channel process 
kicks off at certain neutrino energies, overcoming phase space barrier which renders the cross 
section to non-physical values at lower energies. The incoming UHE neutrino energy where this 
absorption process starts to kick off is called the cutoff of the neutrino spectrum.   
\begin{figure}[H]
\begin{center}
\includegraphics[width=4.3in,height=2.6in,angle=0]{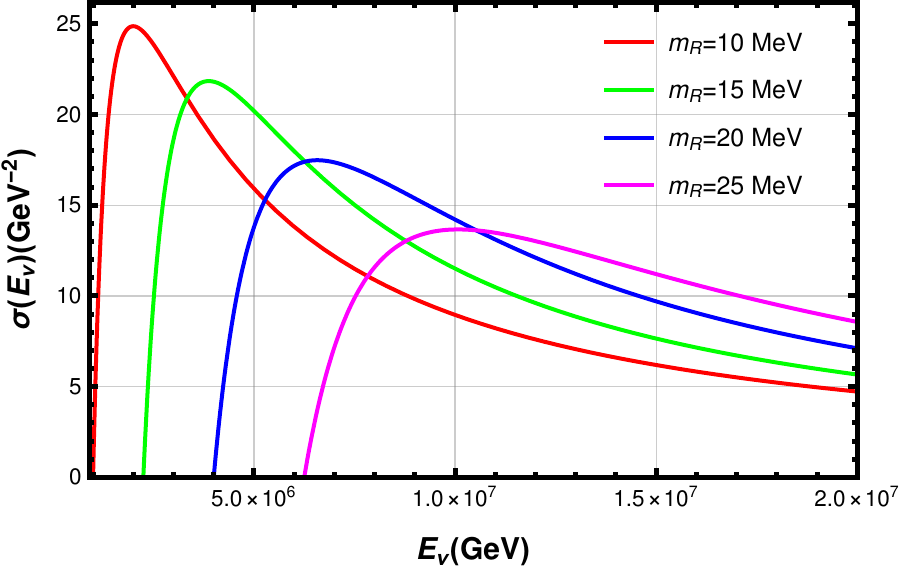}~~
\caption{t-channel absorption cross section and its variation with $m_R$ taking $y \sim 1$ and target neutrino mass 0.1 eV.}
\label{fig:capture}
\end{center}
\end{figure}
The variation of t-channel process cross section with incident neutrino energy is shown in Fig.~\ref{fig:capture}. The cutoff neutrino energy for different $m_R$ values are also shown there. 
Required energy threshold to kick off the process increases with $m_R$ as production of 
heavier particles needs more energy transfer after absorption to enable this 
process kinematically. 
\begin{figure}[H]
\begin{center}
\includegraphics[width=3.2in,height=2.5in,angle=0]{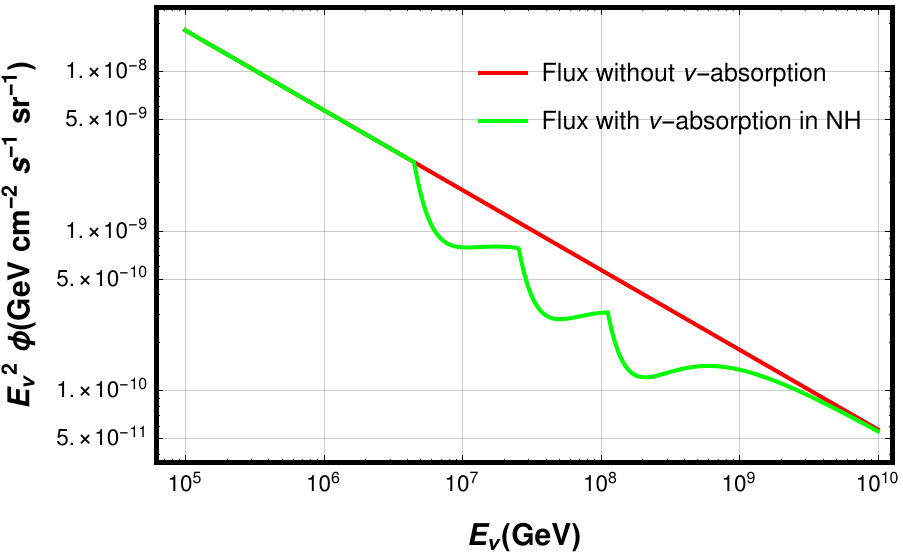}~~
\includegraphics[width=3.2in,height=2.5in,angle=0]{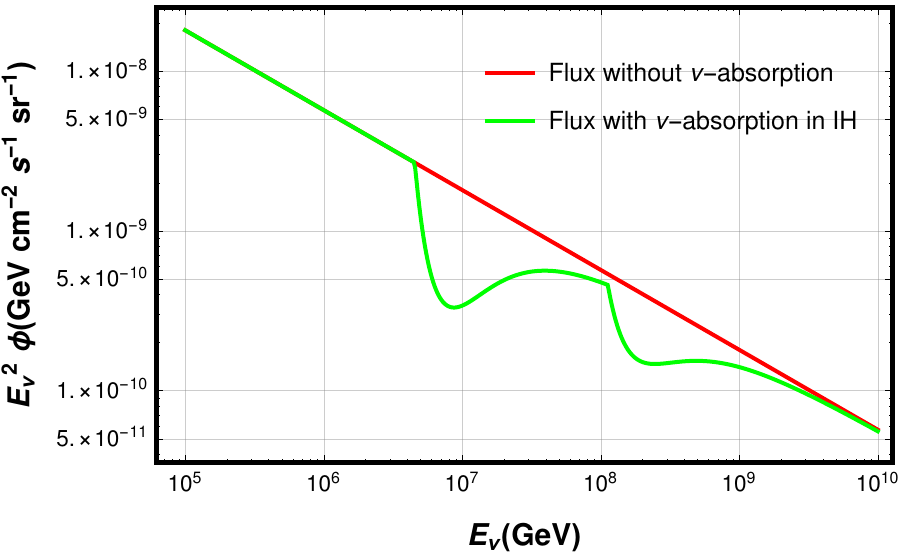}~~
\caption{Comparison of $E^2_{\nu} \times $ flux for the incoming cosmic neutrinos after they got absorbed by the C$\nu$B (green) with that when there is no cosmic neutrino absorption (red) for a. normal mass hierarchy with $(m_1, m_2, m_3) = 2 \times 10^{-3}, 8.8 \times 10^{-3}, 5 \times 10^{-2}$~eV (left) b. inverted mass hierarchy with $(m_1, m_2, m_3) = 4.9 \times 10^{-2}, 5 \times 10^{-2}, 2 \times 10^{-3}$~eV (right)}
\label{fig:ICFluxNuH}
\end{center}
\end{figure}
Here the absorption cross section peaks at t-channel resonant condition, 
i.e., $t = m_h^2$, which in this process at the Lab frame translates to $m_{\nu} E_{\nu}= m_h^2$. 
The cross section will peak at neutrino energies $E_{\nu}$, where this condition will be satisfied. 
The peak will drastically increase the suppression factor for the corresponding neutrino 
energy, as we see in Eq.~\ref{eq:capflux}.

We have fixed two benchmark points in the standard cosmological description of single flux neutrino
propagation. The effects of the presence of the t-channel absorption are shown for those scenarios. 
\begin{itemize}
\item Benchmark Point-I: \\
$\Phi_0 = 1.1 \times 10^{-18} \ (\rm GeV \ cm^2 s \ sr)^{-1}, \ \ \gamma = 2.5$
\item Benchmark Point-II: \\
$\Phi_0 = 1.18 \times 10^{-18} \ (\rm GeV \ cm^2 s \ sr)^{-1}, \ \ \gamma = 2.55 $
\end{itemize} 

\begin{figure}[H]
\begin{center}
\includegraphics[width=3.2in,height=2.5in,angle=0]{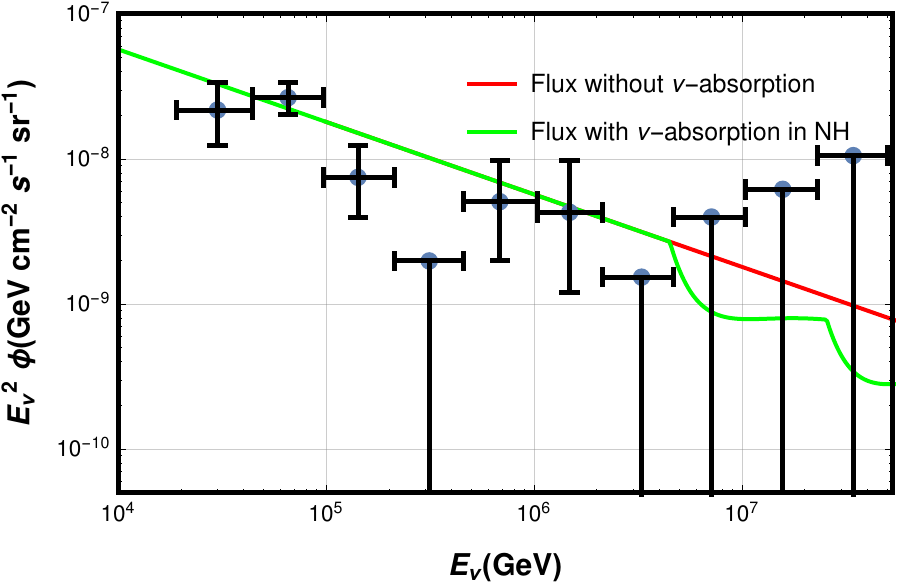}~~
\includegraphics[width=3.2in,height=2.5in,angle=0]{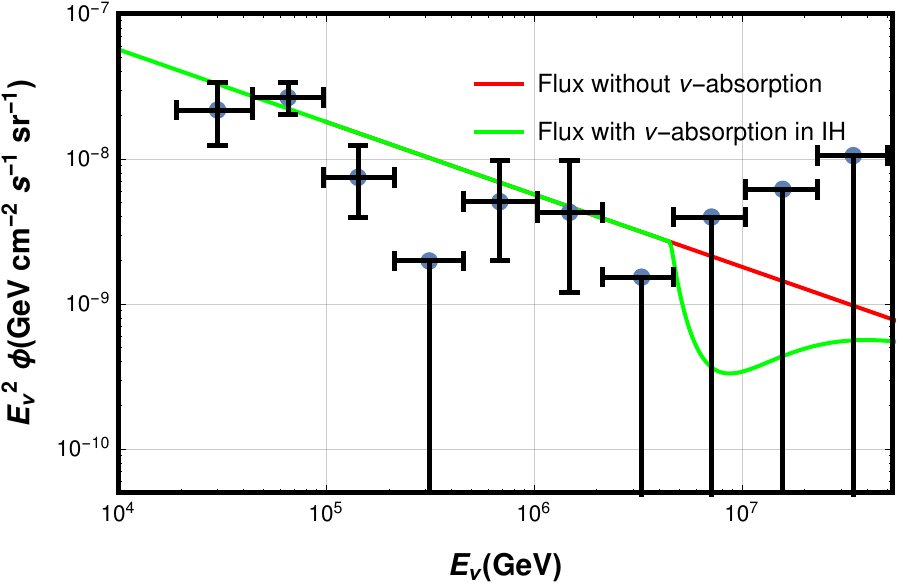}~~
\caption{Comparison of $E^2_{\nu} \times $ flux for the incoming cosmic neutrinos after they got absorbed by the C$\nu$B (green) with that when there is no cosmic neutrino absorption (red) for a. normal mass hierarchy with $(m_1, m_2, m_3) = 2 \times 10^{-3}, 8.8 \times 10^{-3}, 5 \times 10^{-2}$~eV (left) b. inverted mass hierarchy with $(m_1, m_2, m_3) = 4.9 \times 10^{-2}, 5 \times 10^{-2}, 2 \times 10^{-3}$~eV (right). The data points obtained from the IceCube measurement are given in black. Yukawa couplings here are taken to be 0.1 for the representation purpose.}
\label{fig:ICflux}
\end{center}
\end{figure}

We fix the initial UHE neutrino energy cutoff at $4.5$ PeV i.e. the t-channel resonant absorption process will start to contribute only at incident neutrino energies higher that $4.5$ PeV. 
To set this cutoff we need to have one $N_R$ with mass at around $m_R \approx 15$~MeV. 
In Fig.~\ref{fig:ICFluxNuH}, we plot the quantity $E_{\nu}^2 \Phi_{\nu}$ to show how the incoming neutrino flux can be modified due to this t-channel cosmic neutrino absorption. 
This results in multiple dips in neutrino flux spectrum in some particular energies, for both the normal and inverted neutrino mass hierarchies. The first lower energy dip happens at $E_{\nu} \sim 5$~PeV, corresponding to the heaviest neutrino present in the cosmic neutrino background (C$\nu$B). For normal hierarchy the neutrino masses are well separated and therefore three different neutrino mass eigenstates produce three dips in the neutrino flux spectrum. On the other hand, for inverted mass hierarchy with tiny $\Delta m_{12}^2$, two heavier neutrino states have masses $m_1 \approx m_2$. This results in a deeper first dip in the flux, due to combined effect of cosmic neutrino absorption by both $\nu_{1,2}$. How this flux spectrum looks like compared to the the measured flux at IceCube is shown in Fig.~\ref{fig:ICflux}.

\begin{figure}[tbh!]
\begin{center}
\includegraphics[width=3.2in,height=2.5in,angle=0]{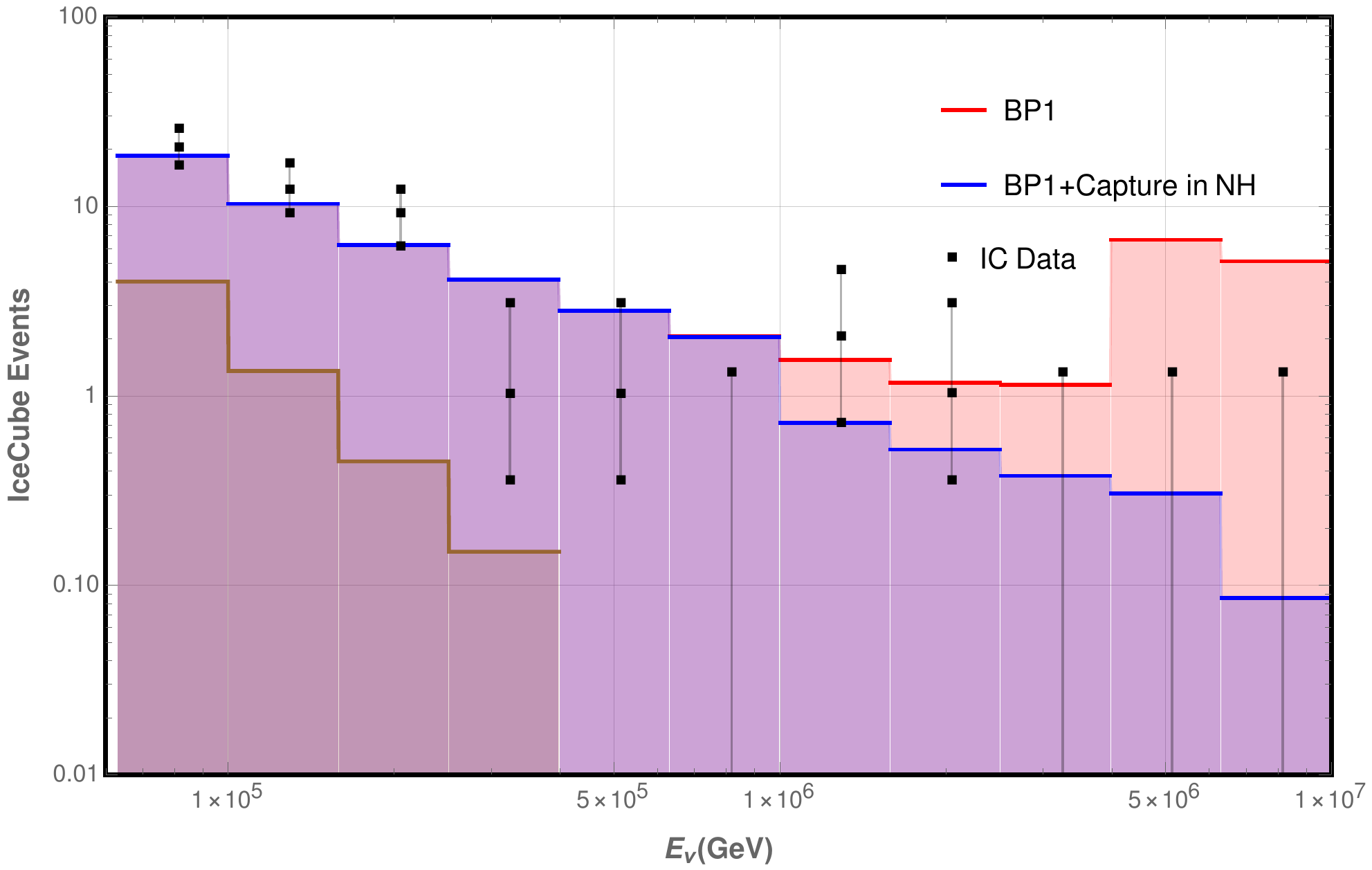}~~
\includegraphics[width=3.2in,height=2.5in,angle=0]{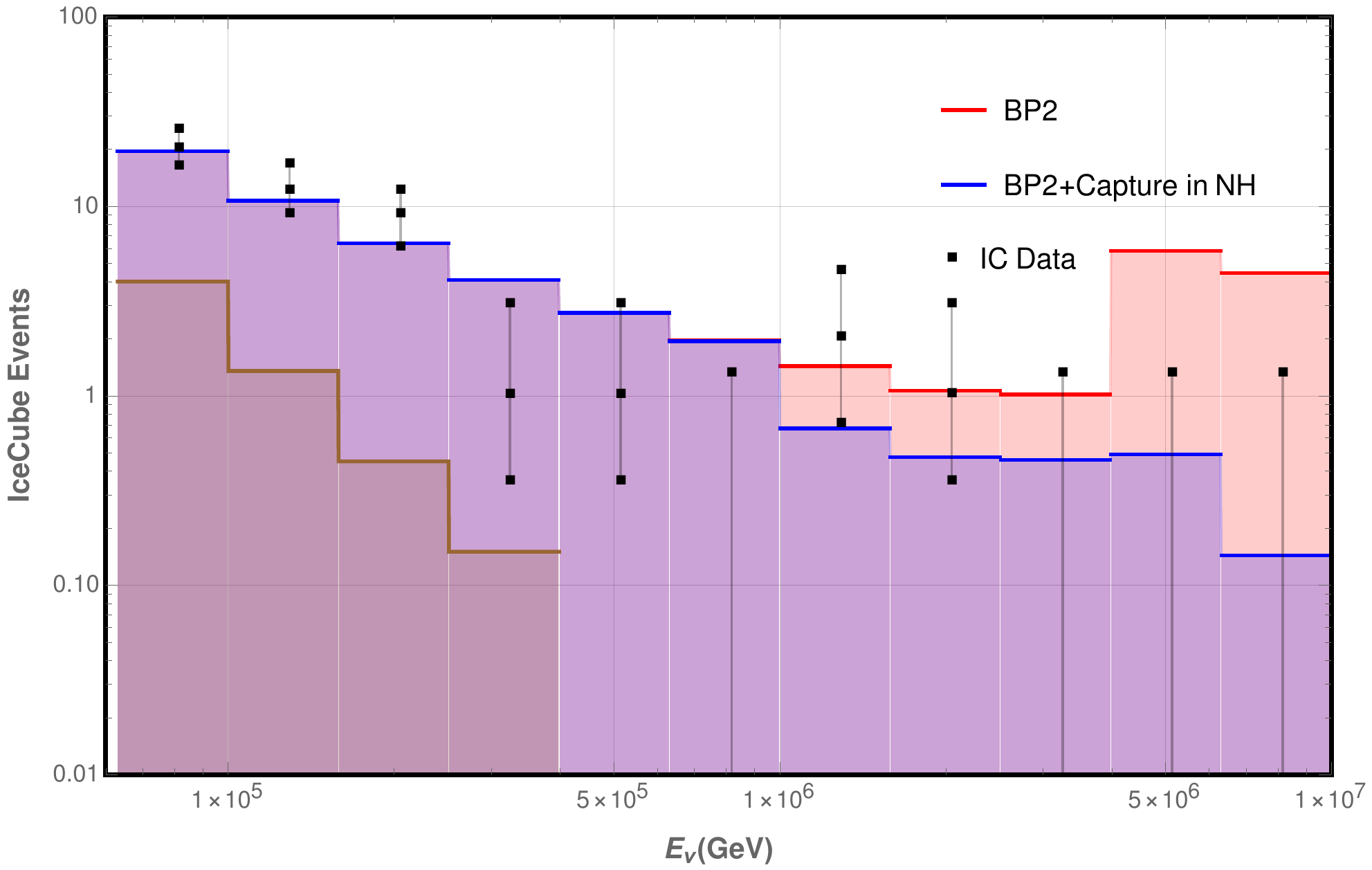}~~ \\ 
\includegraphics[width=3.2in,height=2.5in,angle=0]{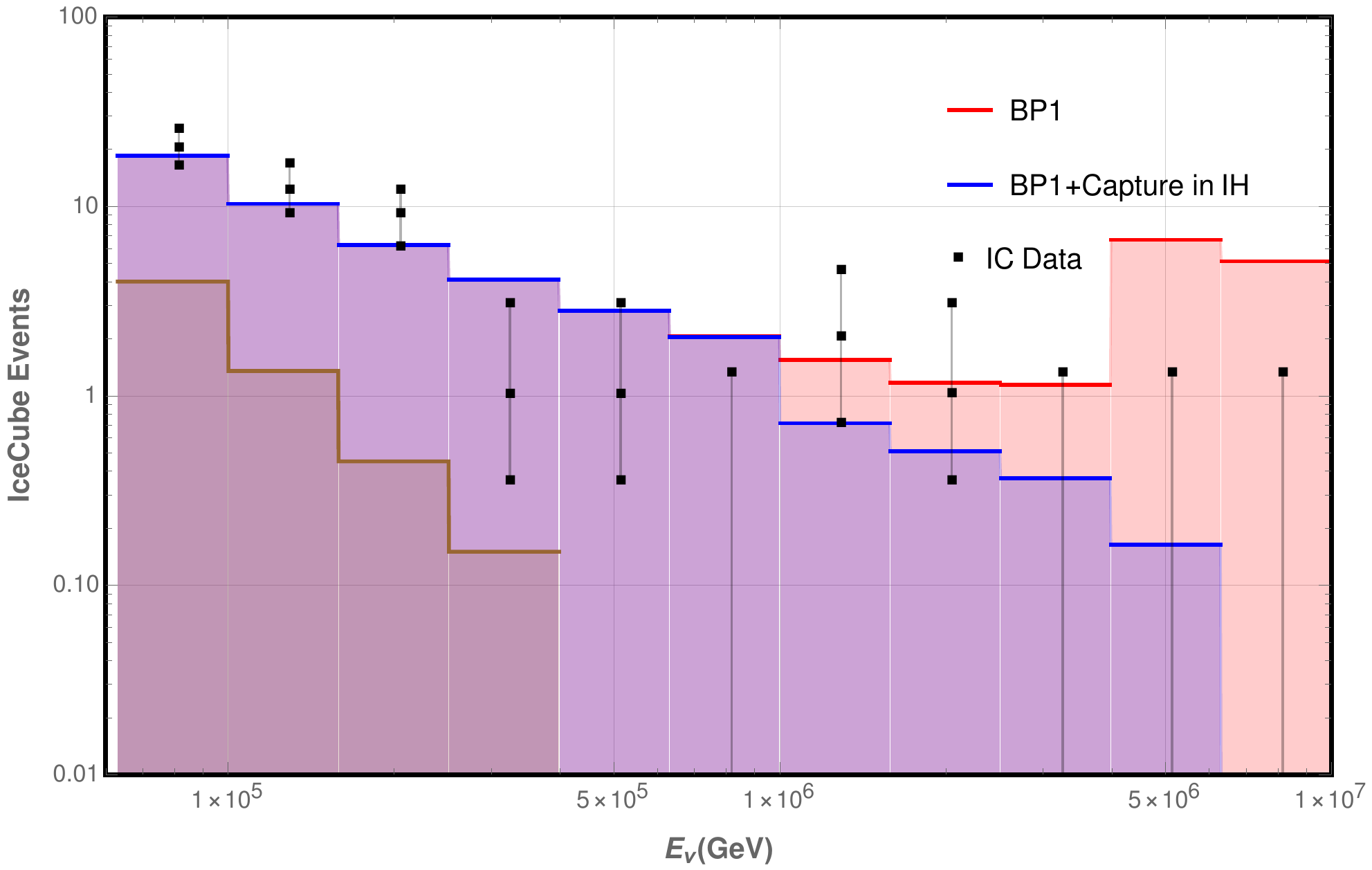}~~
\includegraphics[width=3.2in,height=2.5in,angle=0]{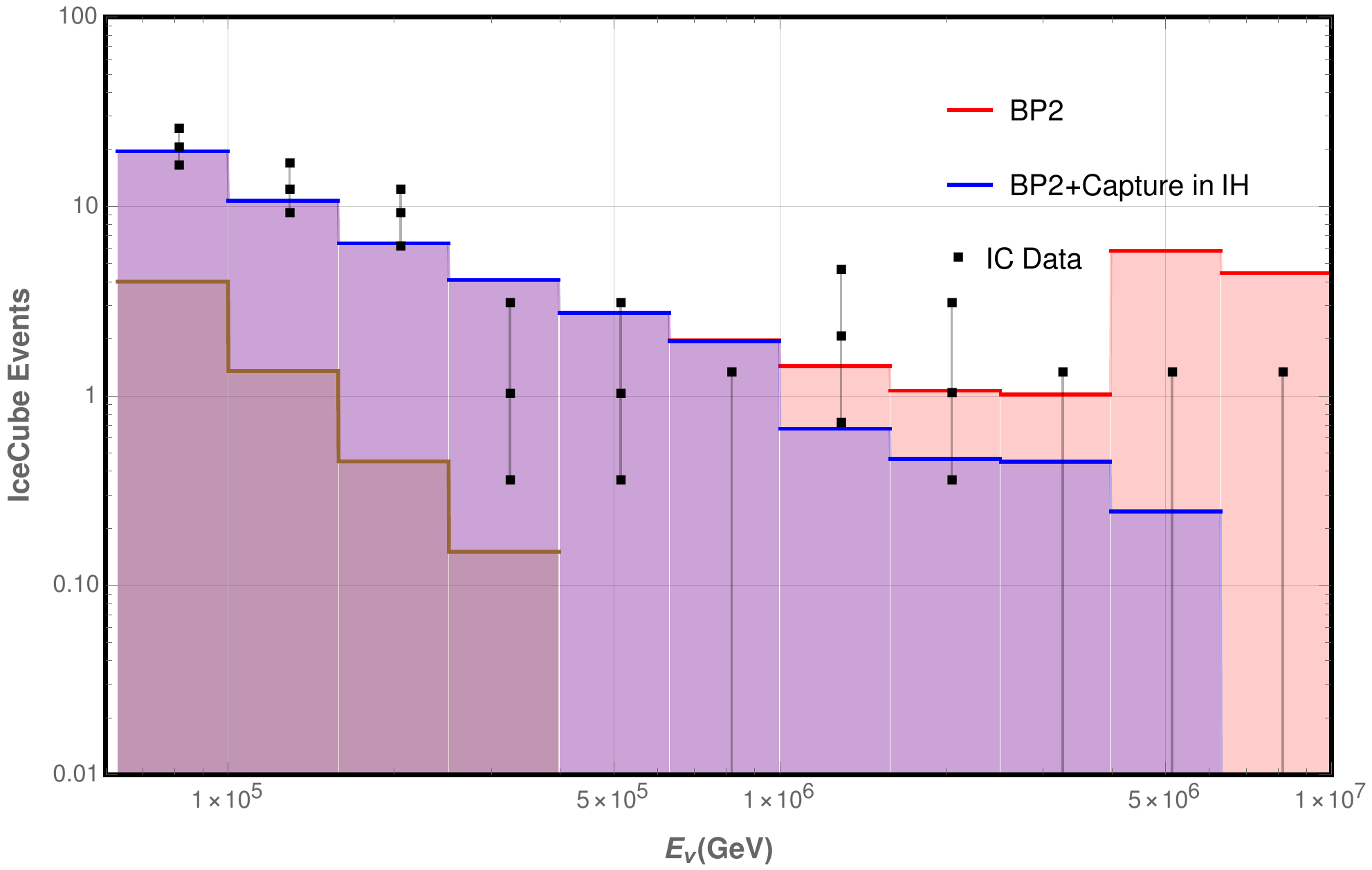}~~ 
\caption{IceCube event spectrum with (violet) and without (red) neutrino capture for Benchamark Point-I (left) and Benchamark Point-II (right). Effect on the event spectrum due to neutrino absorption is shown for normal hierarchy (upper row) and inverted hierarchy (lower row). The atmospheric background is given in brown. Here we have taken $m_R \sim 15$ MeV and $y \sim 1$.}
\label{fig:IC2}
\end{center}
\end{figure}

Effect of the t-channel resonant absorption in this model, resulting in neutrino flux suppression at 
particular energies of IceCube spectrum is shown in Fig.~\ref{fig:IC2}. For the benchmark points mentioned, incoming cosmic neutrinos start interacting with C$\nu$B and get absorbed only when their energy is more than $4.5$ PeV. Those neutrinos therefore do not reach the IceCube detector to deposit energy or leave a track there. Suppression of incoming neutrino flux above $E_{\nu} \ge 4.5$~PeV 
that can possibly interact with the electron in the IceCube results in absence of the 
Glashow resonance at $6.3$~PeV. Amount of neutrino flux suppression at different energies depends
on the t-channel absorption cross section that peaks at an energy determined by the 
light scalar propagator mass ($m_h$) and then decreases with increasing energy. 
Setting $m_h \approx 10$~MeV with a cutoff at $E_{\nu} \ge 4.5$~PeV, fixing the C$\nu$B and $N_R$ masses, we get maximum suppression of neutrino flux at the 11th energy bin ($6.3-10$~PeV) of the IceCube spectrum. 
As the incident neutrino energy does not entirely get deposited at the IceCube, 
few high energetic incident neutrino contribute to the lower energy bins. Here, also due to 
absence of some UHE neutrinos, event count in the nearest energy bins also get a bit suppressed.
Once the t-channel maximal cross section and consequently the sharpest dip occurs in the incident neutrino spectrum, this process cross section starts decreasing with $E_{\nu}$ and eventually matches with the original single power law astrophysical neutrino spectrum at some higher energy. The Yukawa 
coupling also changes the strength of the cross section and therefore can modify the amount of 
suppression we can see in the IceCube spectrum. 
Due to the presence of a cutoff in the t-channel absorption, which we can fix at $E_{\nu} \sim 4.5$~
PeV, this process does not affect the IceCube spectrum at lower energy bins where the we observe 
no neutrino event suppression as shown in Fig.~\ref{fig:IC2}. For the normal hierarchy, the heaviest neutrino mass eigenstate causes the dip at higher energy IceCube bins whereas for the inverted hierarchy two almost degenrate heavier mass eigenstates cause a sharper dip at the same energy bins. 

The best fit single power law flux parametrization is $\phi_0=2.46\pm0.8\times 10^{-18}$ and $\gamma=2.92$ \cite{Williams:2018kpe}, which is different from our benchmark points. We quantify the goodness-of-fit for our chosen benchmark points as well as the IceCube best fit parameters by the respective $\chi^2$ values. We find that although the fitting, in the absence of absorption, worsens for our benchmark points compared to the IceCube best fit but it improves significantly once we include the t-channel resonant absorption in the analysis. The $\chi^2$ value for the IceCube best fit is 21.9 and the same for our benchmark points is shown in table \ref{tab:chi2}.
\begin{table}[h!]
\begin{center}
\begin{tabular}{|c|c|c|c|c|}
\hline
& \multicolumn{2}{c|}{BP1} & \multicolumn{2}{c|}{BP2} \\
\cline{1-5}
{Without Absorption}  & \multicolumn{2}{c|}{49.25} & \multicolumn{2}{c|}{38.55}  \\
\cline{1-5}
\multirow{2}{*}{With Absorption}  & NH & IH & NH & IH \\
\cline{2-5}
 &7.23 &7.17 &6.86 &6.72 \\
\hline
\end{tabular}
\end{center}
\caption{The goodness-of-fit for our all benchmark points is shown here. The numerical entries are the $\chi^2$ values.}
\label{tab:chi2}
\end{table}

\section{Singlet Dark Matter}
\label{sec:singdm}
We extend the model of Sec.~\ref{sec:model} to include a gauge singlet scalar $\chi$ which is odd under $Z_2$ symmetry. The scalar potential is now modified with the addition of
\be
V_{DM}=\frac{1}{2}m_{\chi}^{2}\chi^{2}+\frac{\lambda_{\chi}}{4!} \chi^{4}+\lambda \Phi_{2}^{\dag}\Phi_{2} \chi^{2} 
\ee

\subsection{Relic density}
\begin{figure}[tbh!]
\begin{center}
\includegraphics[width=2.2in,height=1.2in,angle=0]{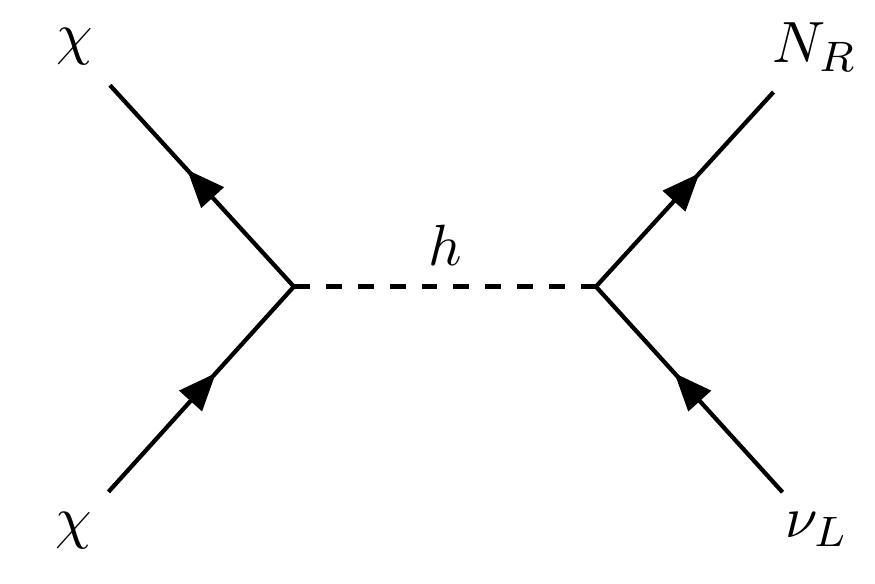}~~
\caption{DM annihilation into neutrinos}\label{fig:relic}
\end{center}
\end{figure}

Based on the structure of the model explained in Sec. \ref{sec:model} DM-DM annihilation to the SM particles through SM Higgs $H$ is kinematically suppressed because of the relatively high Higgs mass. DM can annihilate to SM fermions and neutrinos through the CP even scalar of $\Phi_2$. The annihilation to the SM fermions (except $\nu_L$) is suppressed due to the smallness of $\tan \beta$. As a result, the only process which contributes significantly to the relic density is the annihilation of DM into neutrinos through light mediator $h$ as given in figure \ref{fig:relic}.
The annihilation cross section for this process is given by
\be
\sigma(s)=\frac{\lambda^2 y^2}{8 \pi s}\frac{(s-4 m_{\nu}^{2})^{3/2}}{\sqrt{s-4 m_{\chi}^{2}}}\frac{1}{(s-m_{h}^{2})^2}
\ee
where $s$ is the Mandelstam variable and $\lambda$ is the effective coupling shown in figure \ref{fig:relic}. The relic abundance of DM is calculated as
\be
\Omega h^2=\frac{2.14 \times 10^{9} GeV^{-1}}{\sqrt{g_{*}} M_{Pl}}\frac{1}{J(x_{f})}
\ee
where $M_{Pl} = 1.22 \times 10^{19}$ GeV is the Planck Mass, $g_{*}=106.75$ is the total number of effective relativistic degrees of freedom and $J(x_{f})$ is given as
\be
J(x_{f})=\int_{x_{f}}^{\infty} \frac{<\sigma v>(x)}{x^{2}}dx
\ee
and the thermal averaged cross section $<\sigma v>(x)$ is given as
\be
<\sigma v>(x)=\frac{x}{8 m_{\chi}^{5} K_{2}^{2}(x)}\int_{4 m_{\chi}^{2}}^{\infty} \sigma(s) \times (s-4 m_{\chi}^{2})\sqrt{s}K_{1}(\frac{x \sqrt{s}}{m_{\chi}})ds
\ee

\begin{figure}[tbh!]
\begin{center}
\includegraphics[width=3.2in,height=2.5in,angle=0]{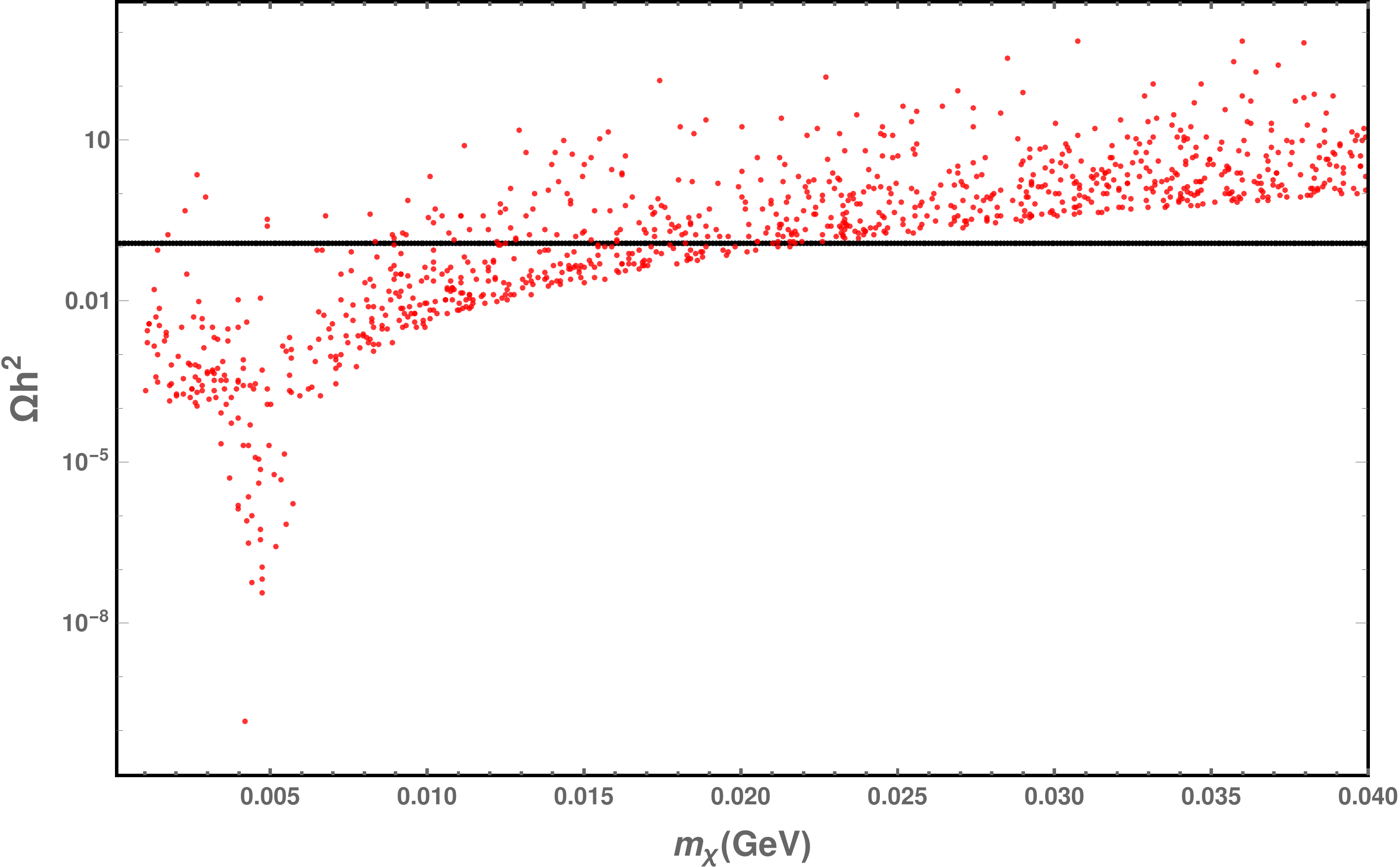}~~
\includegraphics[width=3.2in,height=2.5in,angle=0]{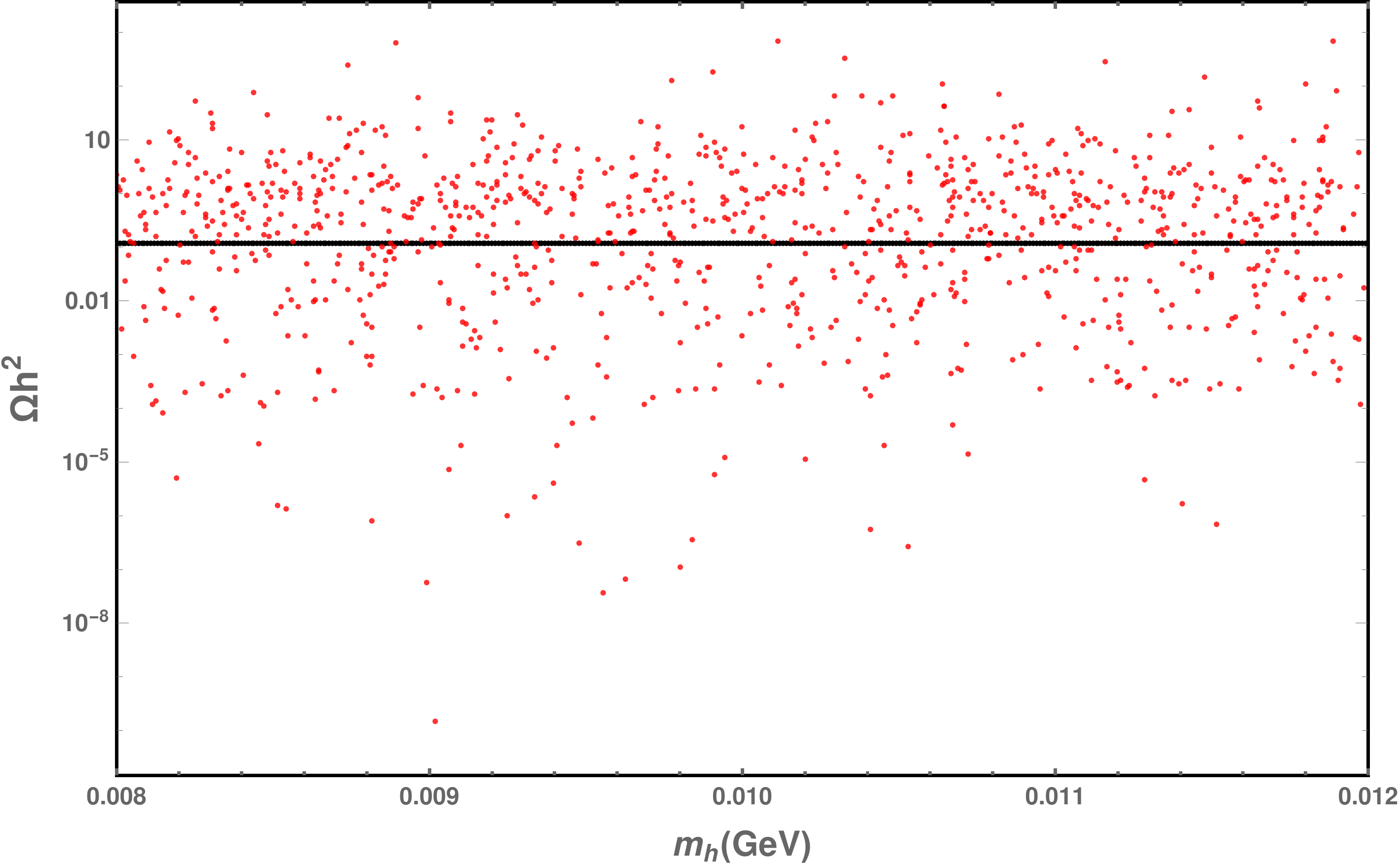}~~
\caption{DM relic density with dark matter mass (left) and mediator mass (right). The black line shows the relic density observed by Planck.}
\label{fig:relicdm}
\end{center}
\end{figure}
where $K_{1}, K_{2}$ are modified Bessel functions, $x=\frac{m_{\chi}}{T}$ where T is the temperature. The $x$ parameter corresponding to the freeze out temperature of DM is analytically given as
\be
x_{f}=\rm{ln}\left(\frac{0.038 g M_{Pl} m_{\chi}<\sigma v>(x_{f})}{(g_{*} x_{f})^{1/2}}\right)
\ee
where $g$ = internal degrees of freedom of DM particle. 
\subsection{Self-interaction of DM}
Despite being extremely successful model for the large scale structure of the Universe, CDM model faces discrepancies at small scales such as, core-cusp problem, missing satellites problem, too-big-to-fail problem and diversity problem. A promising alternative to the CDM scenario, first proposed in ~\cite{spergel} to solve core-cusp problem and missing satellites problem, is self interacting dark matter (SIDM). We refer the reader to ~\cite{tulin} for a detailed review of SIDM and for the solution to the above mentioned problems in the SIDM scenario. The N-body simulations of DM self interaction ~\cite{vogelsberger} suggests DM to have a more Maxwellian distribution as compared to the CDM. Also the presence of self-interaction reduces the density of DM in the central region of halos which results in core~\cite{rocha1} instead of a cusp as it is in the CDM halos~\cite{nfw1997}. As a consequence of self-interaction, SIDM halos are also negligibly elliptical as compared to the CDM halos~\cite{rocha1}. In summary, all these can be understood just by considering the self scattering rate of DM particles,
\be
R_{scat}=\frac{\sigma v_{rel} \rho_{DM}}{m_{\chi}}
\ee
where $\rho_{DM}$ and $v_{rel}$ respectively are the DM density and characteristic relative velocity of DM at a particular scale, with $m_{\chi}$ and $\sigma$ being DM mass and self scattering cross section respectively. Since $\rho_{DM}$ and $v_{rel}$ are different at different scales, the ratio $\sigma/m_{\chi}$ needed to explain the observations would depend on the observation scale. SIDM N-body simulations~\cite{manojprl} predict $\sigma/m_{\chi}\sim 1 cm^2/g$ on galaxy scales and $\sigma/m_{\chi}\sim 0.1 cm^2/g$ on cluster scales.

\begin{figure}[tbh!]
\begin{center}
\includegraphics[width=2.2in,height=1.2in,angle=0]{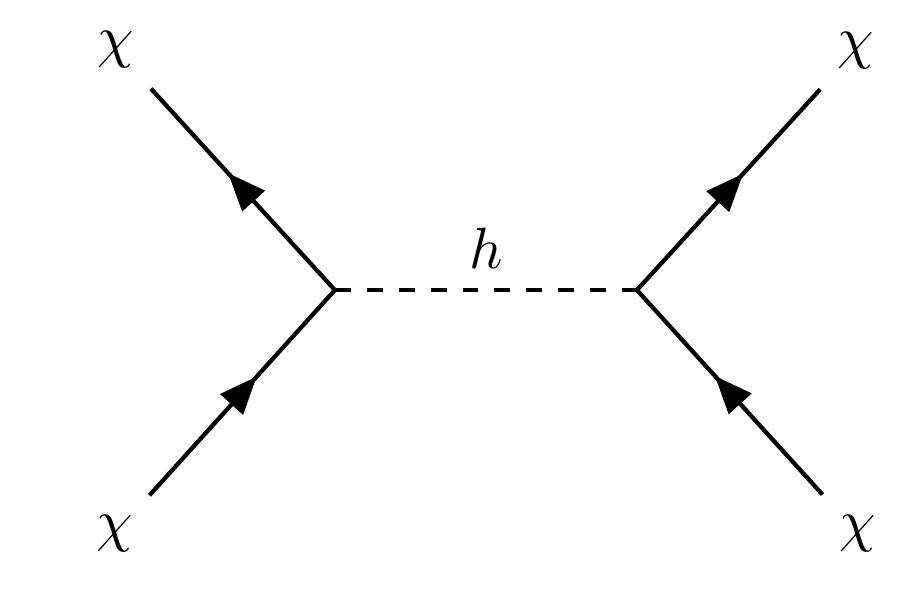}~~
\caption{Self interaction of dark matter}\label{fig:selfint}
\end{center}
\end{figure}

In our model we allow $2 \rightarrow 2$ elastic self scattering of DM particles, which allows the deviations form the CDM predictions. The $2 \rightarrow 2$ DM scattering takes place through $h$ and there is also a 4-point scattering of DM particles. The interaction diagrams are shown in figure \ref{fig:selfint}. 
The ratio $\sigma/m_{\chi}$ is given as
\be
\sigma=\frac{1}{64 \pi m_{\chi}^{2}}\left[\lambda_{\chi}-\frac{(\lambda v_{2})^{2}}{(4 m_\chi^{2}-m_h^{2})}\right]^{2}
\ee

The $\sigma/m_{\chi}$ constraint will be satisfied in this model for allowed parameter space of $\lambda$,$m_h$ and $m_{\chi}$, varying the free parameter $\lambda_{\chi}$.
\section{Discussion and Conclusion}
\label{sec:conc}
A single power law flux of cosmic ray neutrinos cannot fit the event spectrum observed by the IceCube neutrino observatory. The cross section of the Glashow resonance process $\bar{\nu}_{e} e^{-}$ is very large than the neutrino nucleon interaction at energy around 6.3 PeV. Therefore, a dramatic increase in the number of events at this energy is expected, however IceCube has not seen any events at this energy until now. In this work, we discuss a phenomenon that can explain the absence of Glashow resonance at IceCube neutrino detector. We discussed a scenario where cosmic ray neutrinos are absorbed by cosmic neutrino background, therefore causing multiple dips in the cosmic ray neutrino flux, which correspond to three different neutrino mass eigenstates present in cosmic neutrino background, whereas mass information of cosmic ray neutrinos is not significant due to their ultra high energetic nature. Due to different mass splittings, position and depth of the dips are different in normal and inverted mass hierarchies. The occurrence of dips in the neutrino flux is tuned to the energy of the expected GR events, explaining their absence at the IceCube. We use $\nu$2HDM model which allows us to have a t-channel process in which a cosmic ray neutrino interacts with the cosmic neutrino background neutrino through a $\mathcal{O}(10)$ MeV scalar mediator, producing two right handed neutrinos. As a result of this process, dips occur in cosmic ray neutrino flux, of which the lowest energy one happens to cover 5-10 PeV bins. The occurrence of a dip also depends on $m_R$, the masses of the right handed neutrinos being produced and the depth of the dip depends on the mediator mass $m_h$ and the coupling $y$. Fixing $m_R$, $m_h$ and $y$ to the values $m_R=15$ MeV, $m_h=10$ MeV and $y=1$, the energy of the dip in the flux is fixed at the place of GR. This results in the depletion of events at IceCube around 6.3 PeV. The neutrino mass in this process is generated through a very low scale type I seesaw mechanism through the second Higgs doublet. As it turns out, the right handed neutrino mass required to generate the neutrino mass in seesaw is of the order which is needed to explain the missing Glashow resonance at the IceCube. This t-channel neutrino absorption phenomenon can be altered to fit presence of few events in the high energy bins as indicated by more recent IceCube results with nine year data~\cite{Aartsen:2018vtx}. 

\section*{Acknowledgments}
AN and SS are thankful to Namit Mahajan for useful discussions. Authors also thank Ranjan Laha for his suggestions.
\bibliographystyle{JHEP}
\bibliography{nu_Abs.bib}  
  
\end{document}